\documentclass[11pt]{article}
\usepackage{amsmath,amssymb,color}

\usepackage{amsfonts}

\newcommand{\hoch}[1]{$\, ^{#1}$}

\newcommand{\be}{\begin{equation}}
\newcommand{\ee}{\end{equation}}
\newcommand{\bea}{\setlength\arraycolsep{2pt} \begin{eqnarray}}
\newcommand{\eea}{\end{eqnarray}}

\def\ft#1#2{{\textstyle{\frac{\scriptstyle #1}{\scriptstyle #2} } }}
\def\fft#1#2{{\frac{#1}{#2}}}

\def\0{{\sst{(0)}}}
\def\1{{\sst{(1)}}}
\def\2{{\sst{(2)}}}
\def\3{{\sst{(3)}}}
\def\4{{\sst{(4)}}}
\def\5{{\sst{(5)}}}
\def\6{{\sst{(6)}}}
\def\7{{\sst{(7)}}}
\def\8{{\sst{(8)}}}
\def\sst#1{{\scriptscriptstyle #1}}
\def\oneone{\rlap 1\mkern4mu{\rm l}}

\begin{document}

\begin{flushright}
\hfill{CAS-KITPC/ITP-271 \ \ \ \ KIAS-P11036}
\end{flushright}

\vspace{25pt}
\begin{center}
{\large {\bf Killing Spinors for the Bosonic String and the Kaluza-Klein Theory\\
with Scalar Potentials}}

\vspace{10pt}

Haishan Liu\hoch{1}, H. L\"u\hoch{2,3} and Zhao-Long Wang\hoch{4}

\vspace{10pt}

\hoch{1}{\it Zheijiang Institute of Modern Physics\\
Department of Physics, Zheijiang University, Hangzhou 310027, China}

\vspace{10pt}

\hoch{2}{\it China Economics and Management Academy\\
Central University of Finance and Economics, Beijing 100081, China}

\vspace{10pt}

\hoch{3}{\it Institute for Advanced Study, Shenzhen University\\
Nanhai Ave 3688, Shenzhen 518060, China}

\vspace{10pt}

\hoch{4} {\it School of Physics, Korea Institute for Advanced Study,
Seoul 130-722, Korea}

\vspace{40pt}

\underline{ABSTRACT}
\end{center}

The paper consists mainly of two parts. In the first part, we obtain
well-defined Killing spinor equations for the low-energy effective
action of the bosonic string with the conformal anomaly term. We
show that the conformal anomaly term is the only scalar potential
that one can add into the action that is consistent with the Killing
spinor equations.  In the second part, we demonstrate that the
Kaluza-Klein theory can be gauged so that the Killing spinors are
charged under the Kaluza-Klein vector. This gauging process
generates a scalar potential with a maximum that gives rise to an
AdS spacetime. We also construct solutions of these theories.

\vspace{15pt}

\thispagestyle{empty}





\newpage

\section{Introduction}

An underlying property for the successful construction of
supergravities is that the bosonic sector admits consistent Killing
spinor equations, whose projected integrability condition gives rise
to the full set of the bosonic equations of motion.  This property
exists in some intrinsically non-supersymmetric gravity theories as
well.  In fact, the concept of Killing spinor was introduced in
Riemannian Geometry which predates the concept of supersymmetry.  In
pure Einstein gravities in diverse dimensions, which are not
supersymmetric in general, a Killing spinor is defined to satisfy
the Killing spinor equation $D_M \epsilon=0$. The integrability
condition $[D_M, D_N]\epsilon=\ft14 R_{MNPQ}\Gamma^{PQ}\epsilon=0$
suggests that only a subset of Ricci-flat metrics may admit Killing
spinors. However, the projected integrability condition, namely
\begin{equation}
\Gamma^M [D_M, D_N]\epsilon = \ft12 R_{MN}\Gamma^M\epsilon=0\,,
\end{equation}
is satisfied automatically by the Einstein equation of motion.
However, for generic gravity theories with scalar and form fields
that cannot be supersymmetrized to become supergravities, consistent
Killing spinor equations exist rarely.

Killing spinors for domain wall solutions in some scalar-gravity
theories in general dimensions solutions were studied in
\cite{fnss}, where the scalar potentials can be expressed in terms
of superpotentials. We shall review this example and present the
Killing spinor equations in section 2. A far more non-trivial
example has recently been discovered that the low-energy effective
action of the bosonic string in an arbitrary dimension admits
Killing spinor equations \cite{lpw,lw}. Furthermore, this property
is up held even when arbitrary Yang-Mills fields are included, at
least to the $\alpha'$ quadratic curvature order \cite{lw}. This
suggests that the hidden pseudo-supersymmetry associated with the
existence of the Killing spinor equations is a stringy property,
regardless whether the string is supersymmetric or not. It reflects
some generalized geometric property of string theory.

    One clue of constructing Killing spinor equations for the
bosonic string is from the observation of the general study of
pseudo-Killing spinor equations for Einstein gravity coupled to an
$n$-form field strength \cite{lw0}.  By pseudo-Killing spinor
equations, we mean that they are not consistent equations, in that
the projected integrability conditions have extra constraints on the
fields in additional to the equations of motion.\footnote{It should
be emphasized that pseudo-Killing spinors can still be extremely
useful in constructing non-trivial solutions including
non-supersymmetric bubbling and less-bubbling AdS spaces, in which
extra constraints can either be easily satisfied or give
restrictions to the solution space \cite{lw0}. However, in this
paper, we shall be concerned only with theories that admit the
(consistent) Killing spinor equations.} Interestingly, there appears
to be a mysterious connection between the consistency of
Kaluza-Klein sphere reduction of a theory and the existence of
Killing spinor equations. Since it was long established that it is
consistent to perform the $S^3$ and $S^{D-3}$ reductions of the
effective action of the bosonic string \cite{clpred}, Killing spinor
equations were proposed in \cite{lpw} and it was verified that their
projected integrability conditions indeed give rise to the full set
of the equations of motion, without any further constraint. The
result was then generalized to include the $\alpha'$ correction
terms in \cite{lw}.

     However, in the previous discussions \cite{lpw,lw}, the
conformal anomaly term arising from the bosonic string theory in the
non-critical dimensions is set to zero.  In section 3, we construct
Killing spinor equations with this term included.  The construction
can be expected to be successful since this term has been shown not
to spoil the consistency of the sphere reduction \cite{clpred}.  In
fact, we consider a general scalar potential that can be expressed
in terms of a generic superpotential.  Although Killing spinor
equations exist for either the bosonic string without the scalar
potential, or the pure scalar-gravity system, the projected
integrability conditions of the full system give rise to extra
conditions that fix the scalar potential.  It turns out that the
conformal anomaly term and only this scalar potential works with the
Killing spinor equations.  We also find that the system can be
pushed to include the $\alpha'$-order corrections as well.

       Killing spinor equations also exist in the Kaluza-Klein
theory that is the $S^1$ reduction of pure gravity \cite{lpw}.  In
section 4, we consider adding a scalar potential.  We find that as
in the case of the string theory, the projected integrability of the
Killing spinor equations restricts the scalar potential to a single
particular exponential term, analogous to the conformal anomaly term
in string theory.  We compare the results with the $S^1$ reduction
of pure gravity with a cosmological constant in the appendix and
find that these two scalar potentials are different.

       The existence of a vector field in Kaluza-Klein theory
provides a possibility of gauging the theory such that the Killing
spinors are charged under the gauge symmetry.  In section 5, we
demonstrate that the effect of the gauging is that both the
superpotential and the scalar potential contain additional
exponential scalar factors. The (super) potential has a maximum that
gives rise to  the AdS spacetime.  This gauged Kaluza-Klein theory
can be embedded in known gauged supergravities in four, five and
seven dimensions.  We obtain the charged AdS black hole and discuss
its properties.

       One advantage of having Killing spinor equations for a pure
bosonic system is that it may help to construct a general class of
BPS solutions that preserve Killing spinors. Some of these solutions
are unlikely to be found by trying to solve the bosonic equations of
motion alone. In section 6, we use the $D=5$ string theory as an
example to demonstrate this point and obtain the most general
solutions involving Killing spinors.  We conclude the the paper in
section 7.

\section{Scalar-gravity theory}

In supergravities, there are two types of solutions.  Those that
admit Killing spniors are called BPS solutions.  These solutions
preserve certain fractions of the supersymmetry of the theories. The
other type of solutions are non-supersymmetric.  However, it was
observed that even the Schwarzschild-AdS black hole can be solved by
the super-potential method in which the second-order differential
equations can be successfully reduced to a set of first-order
equations {\it via} a super potential \cite{clv,lpv}. (See also
\cite{psrv}.) This suggests that certain non-supersymmetric systems
may exhibit characteristics of supersymmetry; they are {\it
pseudo}-supersymmetric.  Pseudo-supersymmetry for de Sitter
``supergravity" was discussed in \cite{ds1,ds2}. Note that, owing to
the lack of clear definition, the concept of pseudo-supersymmetry in
the past literatures may refer both solutions and abstract theories.

   In this section, We consider gravity coupled to a scalar with a
scalar potential in $D$ dimensions.  The Lagrangian is given by
\begin{equation}
{\cal L}_D = \sqrt{-g} \Big(R - \ft12(\partial\phi)^2 - V(\phi)
\Big)\,.
\end{equation}
We now determine the condition for $V$ such that the above system
admits consistent Killing spinor equations.  Inspired by
supergravities, we introduce a super potential $W(\phi)$ in the
Killing spinor equations, which are given by
\begin{equation}
D_M\eta + \fft{W}{2\sqrt2\,(D-2)}\Gamma_M \eta =0\,,\qquad
\Gamma^M\partial_M\phi \eta -\sqrt2\,W' \eta=0\,,
\end{equation}
where a prime denotes a derivative with respect to $\phi$. In the
above equations, we start with two arbitrary coefficients for the
$W$ and $W'$ terms. The consistency of the projected integrability
condition fixes the above two coefficients. Furthermore, it relates
the potential $V$ to the super potential $W$ as follows
\begin{equation}
V=\Big(\fft{d W}{d\phi}\Big)^2 - \fft{D-1}{2(D-2)}
W^2\,,\label{vwrelation}
\end{equation}
The projected integrability conditions are then given by
\begin{equation}
(R_{MN} - \ft12 \partial_M \phi \partial_N \phi - \fft{V}{D-2}
g_{MN}) \Gamma^M \eta=0\,,\qquad (\Box \phi - V')\eta =0\,.
\end{equation}
Thus provided that the potential $V$ can be expressed in terms of a
superpotential $W$, the scalar-gravity theory has well-defined
Killing spinor equations.  The Killing spinor equations for such a
scalar-gravity theory were studied in the context of AdS domain wall
solutions \cite{fnss}. The application of the scalar-gravity theory
in cosmology was discussed in \cite{sketow}.

\section{Bosonic string with the conformal anomaly term}

The critical dimension for the bosonic string is $D=26$. It suffers
from a conformal anomaly in the dimension $D\ne 26$. It turns out
that the effect of this anomaly is to generate an additional scalar
potential of the dilaton to the effective action \cite{cmpf}
\begin{equation}
{\cal L}_D = \sqrt{-g} \Big(R - \ft12(\partial\phi)^2 - \ft1{12}
e^{a\phi} H_\3^2  - V\Big)\,.\label{bslag}
\end{equation}
where $H_\3=dB_\2$, $a=2\sqrt2/(D-2)$ and
\begin{equation}
V=\ft12(D-26) m^2 e^{-\ft12a\phi}\,.\label{stringV}
\end{equation}
It was shown that it is consistent to perform $S^3$ and $S^{D-3}$
reductions even when the conformal anomaly term is present
\cite{clpred}. In the absence of the conformal anomaly term, the
Killing spinor equations for this system were obtained in
\cite{lpw,lw}. The ansatz for the Killing spinor equations with the
anomaly term is
\begin{eqnarray}
D_M\eta + \ft{1}{96} e^{\fft12a\phi}\Big(a^2 \Gamma_M \Gamma^{NPQ} -
12 \delta_{M}^N\Gamma^{PQ}\Big) H_{NPQ}\,\eta
+ \fft{W}{2\sqrt2\,(D-2)}\Gamma_M \eta &=&0\,,\label{ks1}\\
\Big(\Gamma^M\partial_M \phi+ \ft{1}{12}a e^{\fft12a\phi}
\Gamma^{MNP} H_{MNP}-\sqrt2\,W'\Big)\eta &=&0\,, \label{ks2}
\end{eqnarray}
As we see in section 2, one necessary condition for the above
equation to be consistent is that the scalar potential $V$ and its
superpotential $W$ must be related by (\ref{vwrelation}). Thus,
given that the scalar potential $V$ is known (\ref{stringV}), $W$
can be straightforwardly determined. However, it is still necessary
to check the full projected integrability conditions to verify
whether they are consistent with equations of motion. Furthermore,
one may also ask whether the theory (\ref{bslag}) can admit Killing
spinor equations for other scalar superpotentials.

Thus for now, let us consider that $V$ is given by
(\ref{vwrelation}), but with the superpotential $W$ being
unspecified. We shall use the consistency of the projected
integrability condition to determine the superpotential $W$ and
hence the potential $V$.

     Acting on (\ref{ks2}) with $\Gamma^N D_N$, we find that
\begin{eqnarray}
&&\Big(\nabla^2\phi- \ft1{12}ae^{a\phi}H^2- V'(\phi)\Big)\eta
+\ft1{12}ae^{\fft12a\phi}\Gamma^{NM_1M_2M_3} \nabla_N
H_{M_1M_2M_3}\eta\cr && +\ft14ae^{-\fft12a\phi} \Gamma^{M_2M_3}
\nabla_N\left(e^{a\phi} H^N{}_{M_2M_3}\right)\eta - U_1
e^{\ft12a\phi} \Gamma^{M_1M_2M_3} H_{M_1 M_2 M_3} \eta =0\,,
\label{strintcon1}
\end{eqnarray}
where $V$ is given by (\ref{vwrelation}) and
\begin{equation}
U_1=\fft{1}{3\sqrt{D-2}} (W'' - \fft{1}{\sqrt{2(D-2)}}\, W' -
\fft{1}{D-2} W)\,.
\end{equation}
Thus for the projected integrability condition to be consistent with
the equations of motion, we must have $U_1=0$, implying that
\begin{equation}
W=c_1 e^{-\fft{\phi}{\sqrt{2(D-2)}}} + c_2
e^{\fft{2\phi}{\sqrt{2(D-2)}}} \,.\label{wsol1}
\end{equation}
We now examine the projected integrability condition for
(\ref{ks1}). we find
\begin{eqnarray}
&&\Big[R_{MN}-\ft{1}{2} \partial_M\phi\partial_N\phi
-\ft14e^{a\phi}(H^2_{MN}-\ft{2}{3(D-2)} H^2 g_{MN}) - \fft{V}{D-2}
g_{MN} \Big]\Gamma^{N} \eta \cr 
&&-\ft{1}{6(D-2)} e^{\fft12a\phi}\nabla_N
H_{M_1M_2M_3}\left(\Gamma_{M}\Gamma^{NM_1M_2M_3}
-2(D-2)\delta_M^{[N}\Gamma^{M_1M_2M_3]}\right)\eta \cr 
&&-\ft{1}{2(D-2)}e^{-\fft12a\phi}\nabla_N \left( e^{a\phi}
H^N{}_{M_2M_3}\right)
\left(\Gamma_{M}\Gamma^{M_2M_3}-(D-2)\delta_M^{M_2}
\Gamma^{M_3}\right)\eta\cr 
&&+\fft{1}{4(D-2)^\fft32} U_2 e^{\ft12a\phi} \Big(\Gamma_M
\Gamma^{M_1M_2M_3} - (D-2) \delta^{M_1}_M
\Gamma^{M_2M_3}\Big)H_{M_1M_2M_3} \eta =0\,,\label{strintcon2}
\end{eqnarray}
where
\begin{equation}
U_2=W' + \fft{1}{\sqrt{2(D-2)}} W\,.
\end{equation}
It is easy to see that the solution (\ref{wsol1}) with $c_2=0$
satisfy the equation $U_2=0$.  Thus, the scalar potential is
precisely the conformal anomaly term provided that the non-vanishing
constant $c_2$ is given by
\begin{equation}
c_1=\sqrt{26-D}\,  m\,.
\end{equation}
Thus we demonstrate that the presence of the conformal anomaly term
does not spoil the existence of the consistent Killing spinor
equations. Furthermore, the conformal anomaly term is the only
scalar potential that we can add to the bosonic string such that
Killing spinor equations remain consistent.  The result confirms the
suggestion that there is an underlying generalized geometric
structure associated with Killing spnior equations in string theory,
whether it is supersymmetric or not, critical or non-critical.

In the string frame, defined by $ds^2_{\rm string} = e^{-\fft12
a\phi} ds_{\rm Einstein}^2$, the effective Lagrangian becomes
\begin{equation}
{\cal L} = \sqrt{-g}\, e^{-2\Phi} \Big(R + 4 (\partial\Phi)^2 -
\ft{1}{12} H_\3^2 + \ft12 c_1^2\Big)\,,
\end{equation}
where $\Phi=-\phi/a$.  The Killing spinor equations, after scaling
$\eta\rightarrow e^{-\fft18a\phi} \eta$, become
\begin{equation}
D_M(\omega_-) \eta=0\,,\qquad \Big(\Gamma^M \partial_M \Phi
-\ft{1}{12} \Gamma^{MNP} H_{MNP} + \fft{c_1}{2\sqrt2}\Big) \eta=0\,,
\label{stringks0}
\end{equation}
where $\omega_-$ is the torsionful spin connection, given by
\begin{equation}
\omega_{M\pm}{}^{AB} = \omega_{M}^{AB} \pm \ft12 H_{M}{}^{AB}\,.
\end{equation}
It is interesting to note that the equation $D_M(\omega_-)\eta=0$ is
unmodified by the conformal anomaly.

        It was shown in \cite{lw} that the Killing spinor equations
can be well-defined for the effective action of the bosonic string
with arbitrary Yang-Mills fields up to the $\alpha'$ order. A
crucial property that enables one to write the action up to the
quadratic curvature terms is the first equation in (\ref{stringks0})
\cite{lw}. This equation is unmodified by the conformal anomaly.
This implies that the procedure of obtaining the $\alpha'$-order
Lagrangian works equally well with the presence of the conformal
anomaly. We find that the most general Lagrangian at the tree level,
up to the $\alpha'$ order, is given by
\begin{equation}
{\cal L}_D = \sqrt{-g}e^{-2\Phi}\Big[R+\ft12c_1^2 +
4(\partial\phi)^2 - \ft1{12} H_\3^2 -\ft14\alpha \Big({\rm
tr}'F_\2^2 -
R_{MNAB}(\omega_+)R^{MNAB}(\omega_+)\Big)\Big]\,,\label{alpha'lag}
\end{equation}
where
\begin{equation}
dH_\3 = \ft12\alpha  \Big({\rm tr}(R_\2(\omega_+)\wedge
R_\2(\omega_+))-{\rm tr}' (F_\2\wedge F_\2) \Big)\,.
\label{stringbianchi}
\end{equation}
The Killing spinors equations are given by
\begin{equation}
D_M(\omega_-) \eta=0\,,\quad \Big(\Gamma^M \partial_M \Phi
-\ft{1}{12} \Gamma^{MNP} H_{MNP} + \fft{c_1}{2\sqrt2}\Big)
\eta=0\,,\quad \Gamma^{M_1M_2} F_{M_1M_2}\eta=0\,,\label{stringks}
\end{equation}
We adopt exactly the same notation as in \cite{lw}.

\section{Kaluza-Klein theory with a scalar potential}

In this section we consider Kaluza-Klein theory with a scalar
potential
\begin{equation}
{\cal L}_D= \sqrt{-g} \Big(R - \ft12 (\partial\phi)^2 - \ft14
e^{a\phi} F_\2^2 - V(\phi)\Big)\,,\label{f2lag}
\end{equation}
where $F_\2=dA_\1$, $a=\sqrt{2(D-1)/(D-2)}$.  For $V=0$, the
Lagrangian is the $S^1$ reduction of $(D+1)$-dimensional pure
gravity, and $A_\1$ is the Kaluza-Klein vector.  It was shown in
\cite{lpw} that the system admits well-defined Killing spinor
equations.  Following the discussion of the bosonic string theory
with the conformal anomaly, we now derive the scalar potential $V$
such that consistent Killing spinor equations can still be defined.
As discussed in section 2, a necessary condition is that $V$ can be
expressed in terms of a superpotential $W$ as in (\ref{vwrelation}).
The ansatz for the Killing spinor equations is
\begin{eqnarray}
&&D_M\eta + \fft{\rm i}{8(D-2)}e^{\fft12a\phi}
\Big(\Gamma_M\Gamma^{M_1M_2} - 2(D-2)\delta_{M}^{M_1}
\Gamma^{M_2}\Big) F_{M_1M_2}\eta\cr
&&\qquad\qquad+ \fft{W}{2\sqrt2\,(D-2)}\Gamma_M \eta =0\,,\cr 
&&\Gamma^M\partial_M\phi \eta + \ft{\rm i}{4}ae^{\fft12 a\phi}
\Gamma^{M_1M_2} F_{M_1M_2}\eta {-}\sqrt2\fft{dW}{d\phi}
\eta=0\,.\label{f2kseq1}
\end{eqnarray}
We find that the projected integrability condition for the second
equation in (\ref{f2kseq1}) is given by
\begin{eqnarray}
&&\Big(\nabla^2\phi- \ft1{4}ae^{a\phi}F^2 - V'(\phi)\Big)\eta
+\ft{\rm i}4 ae^{\fft12a\phi}\Gamma^{NM_1M_2} \nabla_N
F_{M_1M_2}\eta\cr && +\ft{\rm i}{2} ae^{-\fft12a\phi} \Gamma^{M_2}
\nabla_N\left(e^{a\phi} F^N{}_{M_2}\right)\eta - \ft{\rm i}2
\sqrt{\ft{D-1}{D-2}}\, U_1e^{\fft12 a\phi} \Gamma^{M_1M_2} F_{M_1
M_2}\eta=0\,,
\end{eqnarray}
where $U_1$ is a function of $W$, $W'$ and $W''$, given by
\begin{equation}
U_1=W'' + \sqrt{\fft{2}{(D-1)(D-2)}}\, W' - \fft{D-3}{2(D-2)} W\,.
\end{equation}
Since the structure $U_1$ is unrelated to any of the equations of
motion, it has to vanish on its own. The solution to $U_1=0$ is
\begin{equation}
W=c_1 e^{\fft{D-3}{\sqrt{2(D-1)(D-2)}}\,\phi}+ c_2
e^{-\fft{D-1}{\sqrt{2(D-1)(D-2)}}\,\phi}\,.\label{wsol2}
\end{equation}
The projected integrability condition for the first equation of
(\ref{f2kseq1}) is given by
\begin{eqnarray}
&&\Big[R_{MN} - \ft12 \partial_M \phi \partial_N\phi - \ft12
e^{a\phi} (F_{MN}^2 - \ft{1}{2(D-2)} F^2 g_{MN}) - \ft1{D-2}V(\phi)
g_{MN}\Big]\Gamma^N \eta \cr
&&-\ft{\rm i}{4(D-2)} e^{\fft12a\phi}\nabla_N F_{M_1M_2}\Big(
\Gamma_M\Gamma^{NM_1M_2} - 3(D-2) \delta_{M}^{[N} \Gamma^{M_1M_2]}
\Big)\eta \cr 
&&-\ft{\rm i}{2(D-2)} e^{-\fft12 a\phi} \nabla_N \Big(e^{a\phi}
F^N{}_{M_2}\Big) \Big(\Gamma_M \Gamma^{M_2} - (D-2)
\delta_M^{M_2}\Big)\eta\cr 
&& +U_2 \delta_M^{M_1}\Gamma^{M_2} e^{\fft12a\phi} F_{M_1M_2} \eta
=0\,,\label{f2ksint2}
\end{eqnarray}
where
\begin{equation}
U_2 = \ft{\rm i}2 \sqrt{\fft{D-1}{D-2}}\, W' - \fft{\rm
i\sqrt2(D-3)}{4(D-2)}\, W\,.
\end{equation}
Again the term $U_2$ is unrelated to any of the equations of motion
and it has to vanish. However, it is easy to see that  $W$ in
(\ref{wsol2}) with $c_2=0$ satisfy the equation $U_2=0$. Thus the
scalar potential $V$ is given by
\begin{equation}
V=-\fft{2c_1^2}{D-1}\, e^{\fft{2(D-3)}{\sqrt{2(D-1)(D-2)}}\,\phi}\,.
\end{equation}
Naively one would expect that the this single exponential scalar
potential has an origin as the cosmological constant in  Einstein
gravity in $D+1$ dimensions, since cosmological Einstein gravity
does accept a well-defined Killing spinor equation. However, as we
show in the appendix, the scalar potential from $S^1$ reduction of
the cosmological constant, given by (\ref{cosmopot}), is different
from this structure. Thus the origin of this scalar potential
remains to be understood.

\section{Gauging the Killing spinors and AdS spacetimes}

With the Kaluza-Klein vector, it is possible that we can gauge the
theory such that fermions are charged under the vector.  In the case
of supergravities, this procedure may turn a supergravity theory to
the gauged supergravity, where a scalar potential is generated. Let
us propose that the Killing spinors are charged under $A_\1$. The
equations are now given by
\begin{eqnarray}
&&(D_M+ b A_M)\eta + \fft{\rm i}{8(D-2)}e^{\fft12a\phi}
\Big(\Gamma_M\Gamma^{M_1M_2} - 2(D-2)\delta_{M}^{M_1}
\Gamma^{M_2}\Big) F_{M_1M_2}\eta\cr
&&\qquad\qquad+ \fft{W}{2\sqrt2\,(D-2)}\Gamma_M \eta=0\,,\cr 
&&\Gamma^M\partial_M\phi \eta + \ft{\rm i}{4}ae^{\fft12 a\phi}
\Gamma^{M_1M_2} F_{M_1M_2}\eta {-}\sqrt2\fft{dW}{d\phi}
\eta=0\,,\label{f2kseq2}
\end{eqnarray}
where the constant $b$ is to be determined.  Note that these
equations (\ref{f2kseq2}) are invariant under the gauge
transformation
\begin{equation}
A_\1 \rightarrow A_\1 + d\Lambda\,,\qquad \eta\rightarrow \eta\,
e^{-b\Lambda}\,.
\end{equation}
The projected integrability condition for the second equation in
(\ref{f2kseq2}) remains the same as the previous ungauged case. It
implies that $W$ is given by (\ref{wsol2}).  Note that we can shift
$\phi$ by constant to adjust the relative coefficients of $c_1$ and
$c_2$. Make a convention that the fixed point occurs at $\phi=0$, we
have
\begin{equation}
 c_1=\fft{D-1}{\sqrt2} g\,,\qquad
 c_2=\fft{D-3}{\sqrt2}g\,.
\end{equation}
We have chosen the convention such that $g^2=1$ corresponds to the
AdS with unit length. The projected integrability condition for the
first equation in (\ref{f2kseq2}) gives the same form as
(\ref{f2ksint2}), but now with $U_2$ given by
\begin{equation}
U_2 = \ft{\rm i}2 \sqrt{\fft{D-1}{D-2}}\, W' - \fft{\rm
i\sqrt2(D-3)}{4(D-2)}\, W +2 b\, e^{-\fft12 a \phi}\,.
\end{equation}
Thus the vanishing of $U_2$ implies that
\begin{equation}
b=-{\rm i} \ft{\sqrt2}4\, c_2 =-\ft{\rm i} 4(D-3)g\,.
\end{equation}
The full consistent Killing spinor equations are now given by
\begin{eqnarray}
&&\Big(D_M-\ft14(D-3)\,{\rm i}\, g A_M\Big) \eta +
\fft{W}{2\sqrt2\,(D-2)}\Gamma_M \eta\cr && 
\qquad + \fft{\rm i}{8(D-2)}e^{\fft12a\phi}
\Big(\Gamma_M\Gamma^{M_1M_2} - 2(D-2)\delta_{M}^{M_1}
\Gamma^{M_2}\Big) F_{M_1M_2}\eta=0\cr
&&\Gamma^M\partial_M\phi \eta + \ft{\rm i}{4}ae^{\fft12 a\phi}
\Gamma^{M_1M_2} F_{M_1M_2}\eta -\sqrt2\fft{dW}{d\phi}
\eta=0\,,\label{f2kseqres}
\end{eqnarray}
where the super potential $W$ is completely determined, given by
\begin{equation}
W=\fft{g}{\sqrt2}\Big((D-3) e^{-\fft{D-1}{\sqrt{2(D-1)(D-2)}}\,
\phi} + (D-1) e^{\fft{D-3}{\sqrt{2(D-1)(D-2)}}\, \phi}\Big)\,.
\end{equation}
The corresponding scalar potential is
\begin{equation}
V=-g^2 (D-1) \Big( (D-3) e^{-\sqrt{\fft{2}{(D-1)(D-2)}}\,\phi} +
e^{\fft{\sqrt2\,(D-3)}{\sqrt{(D-1)(D-2)}}\,\phi}\Big)\,.
\label{f2pot}
\end{equation}
It is clear that this potential has a maximum at $\phi=0$ with
$V(0)=-(D-1)(D-2) g^2$.

     We find that the Lagrangian (\ref{f2lag}) with the scalar
potential (\ref{f2pot}) can be embedded in gauged supergravities in
$D=4,5$ and 7. (See, for example, \cite{tenauthor}.)  In the case of
$D=6$, it may also be possible to embed the theory in the $F(4)$
gauged supergraity \cite{romans} coupled to a vector multiplet
\cite{cglp,clpbubble}.  Thus it may not be surprising that the
Kaluza-Klein theories with this scalar potential in $D=4,5,6$ and 7
admit Killing spinor equations. However it is of great interest to
observe that the theory can admit Killing spinor equations in an
arbitrary dimnension $D$.

We obtain the charged black hole solutions in general dimensions,
for this gauged Kaluza-Klein theory; they are given by
\begin{eqnarray}
ds_D^2&=&-H^{-\fft{D-3}{D-2}} f dt^2 + H^{\fft{1}{D-2}}
\Big(\fft{dr^2}{f} + r^2 d\Omega_{(D-2),k}^2\Big)\,,\cr 
F_\2&=&\sqrt{k} \coth\delta dt\wedge dH^{-1}\,,\qquad \phi =\ft12
a\log H\,,\label{chargedbh}
\end{eqnarray}
where
\begin{equation}
H=1+ \fft{\mu\sinh^2\delta}{k\,r^{D-3}}\,,\qquad f=k -
\fft{\mu}{r^{D-3}} + g^2 r^2 H\,,
\end{equation}
and $k=1,0,-1$ corresponding $d\Omega_{(D-2),k}^2$ being the sphere,
torus and hyperbolic spaces. In the case of $k=0$, we need to scale
$\sinh^2\delta\rightarrow k\,\sinh^2\delta$ before sending $k$ to
zero, and hence $F_\2$ for $k=0$ is given by
\begin{equation}
F_\2 = \fft{1}{\sinh\delta} dt\wedge dH^{-1}\,.
\end{equation}
These solutions in relevant gauged supergravities in $D=4,5,6$ and
$7$ have string and M-theory origins \cite{tenauthor}.

We now examine the thermodynamical properties of the black hole
(\ref{chargedbh}), with $k=1$.  The temperature, entropy, electric
potential and charge are given by
\begin{eqnarray}
T&=&\fft{f'(r)}{4\pi H^{1/2}}\Big|_{r=r_+}\,,\qquad S=\ft14 H^{1/2}
r_+^{D-2} \omega_{D-2}\,,\cr 
\Phi&=&(1-H^{-1}) \coth\delta\,,\qquad Q=\fft{D-3}{32\pi} \mu
\sinh(2\delta) \omega_{D-2}\,.
\end{eqnarray}
The mass of the black hole is given by
\begin{equation}
M=\fft{r_+^{D-3}(1 + g^2 r_+^2)(D-1 + (D-3)\cosh(2\delta))}{32\pi
(1-g^2r_+^2 \sinh^2\delta)}\,.
\end{equation}
The BPS limit is given by setting $\mu\rightarrow 0$ while keeping
$\mu\sinh^2\delta=q$ fixed, for which $H=1 + q/r^{D-3}$ and $f=1 +
g^2 r^2 H$ and $\coth\delta=1$. The Killing spinors exist in this
BPS limit, satisfying the following projection
\begin{equation}
\Big(f^{\fft12} + {\rm i}\, \Gamma^t + g r H^{\ft12}
\Gamma^r\Big)\eta=0\,.
\end{equation}
For $k=0$, the extremal limit implies the vanishing of the 2-form
and the resulting solution is the BPS domain wall.

   To conclude this section, we would like to mention that
the general charged rotating black holes in the Kaluza-Klein theory
with the scalar potential (\ref{f2pot}) was obtained in \cite{wu}.

\section{A class of general ``BPS'' solutions}

The existence of well-defined Killing spinor equations can be a
powerful tool in finding ``BPS'' solutions that preserve at least
one Killing spinor \cite{gghpr}.  As an example, let us consider the
effective action of the bosonic string at $D=5$, with the conformal
anomaly term set to 0.  The Killing spinor equations are given by
(\ref{ks1}) and (\ref{ks2}), with $W=0=W'$. In $D=5$, the spinors
are pseudo-Majorana, and hence all possible bi-spinors take the
following form
\begin{equation}
f={\rm i}\bar\eta\eta\,,\qquad K^M=\bar\eta\Gamma^M\eta\,,\qquad
Y_{(3)}^{MN}=\bar\eta\Gamma^{MN}\eta\,,
\end{equation}
which are real, together with
\begin{equation}
Y^{MN}=\bar\eta^c\Gamma^{MN}\eta\equiv Y_{(1)}^{MN}+i
Y_{(2)}^{MN}\,,
\end{equation}
which is complex. Here, we have $\bar \eta = \eta^\dagger \Gamma_0$
and $\bar \eta^c=\eta^T C$, where $C$ is the charge conjugation
matrix.  We find the following identity for these bi-spinors:
\begin{eqnarray}
\nabla_{M}f&=&\ft1{18} e^{\fft12a\phi}\epsilon_{M} {}^{M_1M_2M_3N}
H_{M_1M_2M_3}K_N\,, \label{5ddf1}
\\
\nabla_{M}K_N&=&\ft1{18}
e^{\fft12a\phi}\left(f\epsilon_{MNM_1M_2M_3}
H^{M_1M_2M_3}+3H_{MNM_3}K^{M_3} \right) \,, \label{5ddf2}
\\
\nabla_{M}Y_{(3)}^{N_1N_2}&=& \ft1{18}
e^{\fft12a\phi}\left[6Y_{(3)}{}_{MM_3}H^{N_1N_2M_3}
-6\delta_M^{[N_2}Y_{(3)}{}_{M_2M_3}H^{N_1]M_2M_3}\right.\cr 
&&\qquad\qquad\left.- 6H_{M}{}^{M_3[N_2}
Y_{(3)}^{N_1]}{}_{M_3}\right]\, \,, \label{5ddf3}
\\
\nabla_{M}Y^{N_1N_2}&=& \ft1{18}
e^{\fft12a\phi}\left[6Y_{MM_3}H^{N_1N_2M_3}-
6\delta_M^{[N_2}Y_{M_2M_3}H^{N_1]M_2M_3}\right.\cr 
&&\qquad\qquad \left. - 6H_{M}{}^{M_3[N_2} Y^{N_1]}{}_{M_3}\right]\,
\,. \label{5ddf4}
\end{eqnarray}
It follows that $K\equiv K^{M}\partial_M$ is a Killing vector and
\begin{equation}
\nabla_{[M}Y_{(i)N_1N_2]}= 0\,\,,\qquad \nabla_{M}Y_{(i)}^{MN_2}=
\ft13 e^{\fft12a\phi}H_{(i)}{}_{M_2M_3}F^{N_2M_2M_3} \,.
\end{equation}
In addition, we have
\begin{eqnarray}
K^{M}\partial_M \phi=Y_{(i)}\wedge F=0 \label{5ddf5} \,,\qquad
Y_{(i)}\equiv Y_{(i)MN}dx^M\wedge dx^N\,.
\end{eqnarray}
With some lengthy algebra, we also obtain the following product
relations
\begin{eqnarray}
K_M K^M &=& -f^2\,,\qquad K_{N}Y_{(i)}^{NM}=0\,,\\
K_{[M_1}Y_{(i)M_2M_3]}&=&-\ft{1}{6}\epsilon_{M_1M_2N_1N_2N_3}f
Y_{(i)}^{M_1M_2}\label{5dky2}\,,\\
Y_{(i)N_1}{}^{N_3}Y_{(j)N_2N_3}&=& \delta_{ij}\left(f^2g_{N_1N_2}
+K_{N_1}K_{N_2}\right)+\epsilon_{ij}{}^{k}f
Y_{(k)N_1N_2}\label{5dyy1}\,,\\
Y_{(i)[N_1N_2}Y_{(j)N_3N_4]}&=&\ft13\delta_{ij}
\epsilon_{MN_1N_2N_3N_4}fK^M\label{5dyy2}\,.
\end{eqnarray}

    We are now in the position to derive the explicit solutions.
For the case $f\ne 0$, the Killing vector $K$ is time-like and we
can choose the coordinate such that $K=\partial_t$. Without loss of
generality we consider $f>0$.  We find that the general BPS solution
is simply given by
\begin{equation}
ds_5^2=-f^2 dt^2 + f^{-1} ds_4^2\,,\qquad H_\3={*_4
df^{-3}}\,,\qquad e^{a\phi} = f^4\,,
\end{equation}
where the bi-spinor relations define the hyper-K\"ahler structure
for $ds_4^2$. The Bianchi identity and the equation of motion for
the 3-form $H_\3$ further imply that $f^{-3}$ is the harmonic
function on the hyper-K\"ahler space $ds_4^2$.

When $f=0$, the Killing vector $K$ is null and we can choose the
coordinate such that $K=\partial_{v}$. The bi-spinor relations and
the Killing spinor equations imply that the general BPS solution is
given by
\begin{eqnarray}
&&ds^2=H^{-1}(\lambda du^2+2dudv)+H^2(dx_i+A_i du)(dx^i+A^i du)\,,
\cr&&H_\3=-3H^{-4}\partial_iHdx^i\wedge du\wedge
dv-H^{-2}\partial_{i}A_{ j}dx^i\wedge dx^j\wedge du \,,
\cr&&e^{a\phi}=H^4\,.
\end{eqnarray}
Note that there should be no confusion between the the 3-form $H_\3$
and the function $H$. After imposing the Bianchi identity and the
equation of motion for $H_\3$ as well as the Einstein equation along
the $uu$ direction, we find
\begin{eqnarray}
\partial_i\partial^i H^3=0\,,\quad \partial_i\partial^iA_{j} =0\,,\quad
\partial_{[i}H\partial_{j}A_{ k]}=0\,,\quad
\partial_i\partial^i\lambda =-2H^3\partial_{i}A_{ j}\partial^{i}A^{ j}
\,,
\end{eqnarray}
where we have imposed the gauge condition $\partial_u H^3 =
\partial_j (H^3 A^j)$ on $A_j$.

\section{Conclusions}

   In this paper we extend the construction of the Killing spinors
in \cite{lpw,lw} for non supersymmetric theories, by introducing
scalar potentials.  In the case of the effective action of the
bosonic string, we find that the existence of consistent Killing
spinor equations requires that the scalar potential is precisely the
conformal anomaly term, which is a single exponential of the dilaton
field. We obtain the most general tree-level action for the bosonic
string up to the $\alpha'$ order, suggesting that the hidden
pseudo-supersymmetry associated with the Killing spinor equations is
an intrinsic stringy property, regardless whether the theory is
supersymmetric or not.

      We also consider the Killing spinor equations for the
Kaluza-Klein theories with scalar potentials in general dimensions.
If the Killing spinor is neutral under the Kaluza-Klein vector, the
scalar potential also has to be a single exponential term, which has
no fixed point. However, we can gauge the theory such that the
Killing spinors are charged under the $U(1)$ Kaluza-Klein vector,
and the resulting scalar potential has one maximum, giving rise to
an AdS vacuum. We obtain the charged black hole and discuss its BPS
limit. The gauegd Kaluza-Klein theory has obvious application in the
AdS/CFT correspondence.  For $D=4,5,6$ and 7, these theories can be
embedded in gauged supergravities and lifted to higher dimensional
fundamental theories, such as M-theory or the type IIB theory.
However, the origin of theories in general dimensions are not clear.
The existence of Killing spinor equations warrants further
investigations in these gauged Kaluza-Klein AdS theories.

      Although in scalar-gravity theories, Killing spinor equations
can be defined for an arbitrary superpotential, the situation is
much more restrictive when form fields are involved.  So far, the
bosonic string and the Kaluza-Klein theory (\ref{f2lag}) with the
scalar potential (\ref{f2pot}) are the only known non-trivial
examples of intrinsically non-supersymmetry theories that admit
Killing spinor equations. It is of great interest to investigate
systematically the conditions for which Killing spinor equations can
arise.

      Finally we used the Killing spinor equations to obtain the
most general solutions that admits at least one Killing spinor for
the effective action for the five-dimensional bosonic string.

\section*{Acknowledgement}

We are grateful to Chris Pope for useful discussions. H.~Liu is
grateful to KITPC, Beijing, for hospitality during the course of
this work, and is supported in part by the National Science
Foundation of China (10425525,10875103), National Basic Research
Program of China (2010CB833000) and Zhejiang University Group
Funding (2009QNA3015).

\appendix
\section{KK reduction of Killing spinor equations}

Let us consider $(D+1)$-dimensional Einstein gravity
\begin{equation}
{\cal L}_{D+1} = \sqrt{-\hat g} \hat R\,.
\end{equation}
It's Killing spinor is defined by
\begin{equation}
D_M \hat \epsilon\equiv \partial_M \hat \epsilon+ \ft14
\omega^{AB}{}_{M} \hat \Gamma_{AB} \hat \epsilon=0\,.
\end{equation}
Performing the $S^1$ reduction,
\begin{eqnarray}
ds_{D+1}^2 &=& e^{2\alpha\phi} ds_{D}^2 + e^{2\beta\phi} (dz +
A)^2\,,\cr 
\beta &=& - (D-2)\alpha\,,\qquad \alpha^2 = \fft{1}{2(D-1)(D-2)}\,,
\label{s1red}
\end{eqnarray}
the lower-dimensional Lagrangian is
\begin{equation}
{\cal L}_D = \sqrt{-g} (R - \ft12(\partial\phi)^2 - \ft14 e^{a\phi}
F_\2^2)\,,
\end{equation}
where $a=-2(D-1)\alpha$.

      A convenient choice for the vielbein for the metric in
(\ref{s1red}) is given by
\begin{eqnarray}
\hat e^{A}{}_M = e^{\alpha \phi} e^{A}{}_M\,,&& \hat
e^{Z}{}_M=e^{\beta\phi} A_M\,,\cr 
\hat e^{A}{}_z=0\,, && \hat e^{Z}{}_z=e^{\beta\phi}\,.
\end{eqnarray}
Note that $Z$ denotes the flat index.  The spin connection is given
by
\begin{eqnarray}
\hat \omega^{AB} &=& \omega^{AB} + \alpha
e^{-\alpha\phi}(\partial^B\phi \hat e^{A} - \partial^A\phi \hat
e^{B}) - \ft12 F^{AB} e^{(\beta-2\alpha)\phi} \hat e^Z\,,\cr 
\hat \omega^{AZ}&=& -\beta e^{-\alpha\phi}\partial^A\phi \hat e^{Z}
- \ft12 F^A{}_B e^{(\beta-2\alpha)\phi} \hat e^B\,.
\end{eqnarray}

Let us first consider the case when $D$ is even.  The
$(D+1)$-dimensional gamma matrices can be decomposed as
\begin{equation}
\hat \Gamma_A= \Gamma_A\,,\qquad \hat \Gamma_Z = \gamma
\end{equation}
where $\gamma$, with $\gamma^2=1$, is the chiral operator that is
anti-commuting with all gamma matrices.  The Killing spinor is given
by
\begin{equation}
\hat \epsilon=e^{\fft12\alpha\phi}\, \eta\,.
\end{equation}
The reduction of $D_z\hat \epsilon=0$ gives rise to
\begin{equation}
-\gamma \Gamma^M \partial_M \phi \eta + \ft1{4\beta}
e^{(\beta-\alpha)\phi} \Gamma^{MN} F_{MN} \eta=0\,.
\end{equation}
The reduction of $D_M \hat \epsilon=0$ gives rise to
\begin{equation}
D_M\eta + \fft{\gamma}{8(D-2)} e^{(\beta-\alpha)\phi} \Big( \Gamma_M
\Gamma^{M_1 M_2} -2(D-2) \delta_M^{M_1} \Gamma^{M_2}\Big)
F_{M_1M_2}\eta=0\,.
\end{equation}

   We now consider adding a cosmological constant in $D+1$ dimensions,
namely ${\cal L}_{\rm cosmo}=-(D-1)(D-2)\lambda^2 \sqrt{-g}$, which
generates a scalar potential in $D$ dimensions:
\begin{equation}
{\cal L}_{\rm pot} = -(D-1)(D-2)\lambda^2
e^{2\alpha\phi}\,.\label{cosmopot}
\end{equation}
The Killing spinor equation in $D+1$ is modified as
\begin{equation}
\hat D_M \hat \epsilon + \ft12\lambda\hat \Gamma_M \hat
\epsilon=0\,.
\end{equation}
The reduced Killing spinor equations become
\begin{equation}
-\gamma \Gamma^M \partial_M \phi \eta + \ft1{4\beta}
e^{(\beta-\alpha)\phi} \Gamma^{MN} F_{MN} \eta -
\fft{\lambda}{\beta} e^{\alpha \phi} \gamma \eta=0\,.
\end{equation}
and
\begin{eqnarray}
&&D_M\eta + \fft{\gamma}{8(D-2)} e^{(\beta-\alpha)\phi} \Big(
\Gamma_M \Gamma^{M_1 M_2} -2(D-2) \delta_M^{M_1} \Gamma^{M_2}\Big)
F_{M_1M_2}\,\eta \cr 
&&\qquad\qquad + \ft{D-1}{2(D-2)}\lambda e^{\alpha\phi} \Gamma_M
\eta =0\,.
\end{eqnarray}
This implies that
\begin{equation}
W=\sqrt{2}\,(D-1)\lambda\, e^{\alpha\phi}\,.\label{wform}
\end{equation}
Note that this scalar potential is different from given in section
4.

    We now consider the case when $D$ is odd.  The gamma matrix
decomposition is given by
\begin{equation}
\hat \Gamma_M = \sigma_1\otimes \Gamma_M\,,\qquad \hat \Gamma_Z =
\sigma_2\otimes \oneone\,.
\end{equation}
The reduced Killing spinor equations become
\begin{eqnarray}
&&D_M\eta - \fft{\rm i}{8(D-2)}e^{\fft12a\phi}
\sigma_3\Big(\Gamma_M\Gamma^{M_1M_2} - 2(D-2)\delta_{M}^{M_1}
\Gamma^{M_2}\Big) F_{M_1M_2}\eta \cr 
&&\qquad+ \fft{W}{2\sqrt2\,(D-2)}\sigma_1\Gamma_M \eta =0\,,\cr 
&&\Gamma^M\partial_M\phi \eta - \ft{\rm i}{4}ae^{\fft12 a\phi}
\sigma_3 \Gamma^{M_1M_2} F_{M_1M_2}\eta
-\sqrt2\fft{dW}{d\phi}\sigma_1 \eta=0\,,
\end{eqnarray}
where $W$ takes the same form as that in (\ref{wform}). Note that
the Pauli matrices tensor product with the gamma matrices.  The
appearance of these pauli matrices reflects that the Killing spinors
are symplectic Majorana. Not also that since $D$ is odd, so $(D+1)$
is even. This implies that in $(D+1)$ dimensions, the Killing spinor
equation can have an alternative form $\hat D_M \hat\epsilon + {\rm
i}\hat \gamma\hat \Gamma_M\hat \epsilon=0$.  Since $\hat \gamma \sim
\sigma_3$, thus it is equivalent to set $\sigma_1$ in the above
equation to $\sigma_2$.

      The reason we present the Kaluza-Klein reduction of pure
gravity and its Killing spinor equations is to make comparison to
the discussion in section 4.  The abstract construction of the
Killing spinors in the Kaluza-Klein theory is somewhat different
from the dimensional reduction of the Killing spinors in one
dimension higher.  The integrability conditions suggest that they
are equivalent.  However, when the scalar potentials are added, they
becomes inequivalent, each admits a different scalar potential.  In
particular, the Killing spinor equations obtained from the
Kaluza-Klein reduction cannot be gauged and $W$ cannot be augmented
with an extra term to generate AdS spacetimes.  This leaves the
origin of the scalar potential (\ref{f2pot}) intriguing to
investigate.

\end{document}